\documentclass[aps,prc,twocolumn,superscriptaddress,showpacs]{revtex4}
\usepackage{graphicx,color,amsmath,amssymb,bm}

\newcommand{\eq}[1]{Eq.~(\ref{#1})}
\newcommand{\be}{\begin{equation}}
\newcommand{\ee}{\end{equation}}
\newcommand{\bea}{\begin{eqnarray}}
\newcommand{\eea}{\end{eqnarray}}
\newcommand{\bfr}{{\bf r}}
\newcommand{\GV}{G_V}

\newcommand{\fm}{\, \text{fm}}

\newcommand{\mev}{\, \text{MeV}}

\begin{document}

\preprint{NT@UW-08-08}

\title{Isospin-symmetry-breaking corrections to superallowed Fermi 
$\beta$ decay:\\ Formalism and schematic models}

\author{G.A.~Miller}
\email[E-mail:~]{miller@phys.washington.edu}
\affiliation{Department of Physics, University of Washington,
Seattle, WA 98195-1560}
\author{A.~Schwenk}
\email[E-mail:~]{schwenk@triumf.ca}
\affiliation{TRIUMF, 4004 Wesbrook Mall, Vancouver, BC, Canada, V6T 2A3}


\begin{abstract} 
We study the formalism to include isospin-symmetry-breaking 
corrections when extracting the up-down Cabibbo-Kobayashi-Maskawa
matrix element from superallowed $0^{+} \rightarrow 0^{+}$ nuclear 
$\beta$ decay. We show that there are no first order 
isospin-symmetry-breaking corrections to the relevant nuclear matrix
elements. We find corrections to the treatment of Towner and Hardy,
and assess these using schematic models of increasing complexity.
\end{abstract} 


\maketitle

\section{Introduction}

Superallowed Fermi $\beta$ decay provides
the most stringent test of the conserved-vector-current (CVC) 
hypothesis, the most precise value for the up-down 
Cabibbo-Kobayashi-Maskawa (CKM) matrix element $V_{ud}$, and the best
limit on the presence of scalar interactions. With the confirmation 
of CVC, $V_{ud}$ can be extracted with great precision to test the
Standard Model~\cite{HT05,HT05a,Ha07}. For this, one needs to 
evaluate $\sim 1\%$ theoretical corrections that arise due to 
nucleus-dependent isospin-symmetry-breaking (ISB) effects between the parent
and daughter states, and due to radiative effects~\cite{TH02,OB95}.
These corrections are small, but significant, and their associated
theoretical errors at present dominate the uncertainty of
$V_{ud}$ because of the very high precision reached 
experimentally~\cite{TH08}.

In the 2005 survey of Hardy and Towner~\cite{HT05}, the results for
the set of superallowed $0^{+}$~$\rightarrow$~$0^{+}$ transitions
were statistically consistent, after including these theoretical
corrections. However, Penning-trap measurements
of the transition energy for $^{46}$V~\cite{Sa05,Er06b} moved this
case to more than two standard deviations away from the 2005 survey.
This lead Towner and Hardy (TH)~\cite{TH08} to reexamine their
treatment of ISB corrections and to include
the contribution from core orbitals. The latter were found to be
especially important for $^{46}$V and this anomaly disappeared.

In this paper, we study the formalism to include 
ISB corrections, and contrast the TH treatment
to exact results. Before proceeding, we review the necessary
theoretical background, following the discussion in TH~\cite{TH08}.

Superallowed $0^{+}$~$\rightarrow$~$0^{+}$ Fermi $\beta$ decay
depends only on the vector part of weak interactions, and with
CVC, the decay transition ``$ft$ value'' should be independent
of the nucleus:
\be
ft = \frac{2 \pi^3 \hbar^7 \ln 2}{| M_F |^2 \, \GV^2  m_e^5 c^4} 
= {\rm const.} \,,
\label{ftconst}
\ee
where $\GV $ is the vector coupling constant and $M_F$ is the 
Fermi matrix element. CVC depends on the assumption of isospin
symmetry, which is not exact in nuclei, but broken by 
electromagnetic and quark mass effects.
As a result, $M_F$ is reduced from its symmetry value of 
$M_0 = \sqrt{2}$ for $T=1$ parent and daughter states. Following
TH, we introduce the ISB corrections
$\delta_C$ to the Fermi matrix element by
\be
|M_F|^2 = |M_0|^2 \, ( 1 - \delta_C ) \,.
\label{MF}
\ee
In addition, there are radiative corrections to Eq.~(\ref{ftconst}),
but we focus on $\delta_C$ here. These isospin corrections are 
$\sim 1\%$, but must be calculated with a theoretical uncertainty
of $10\%$, to guarantee a desired accuracy of $0.1\%$. This presents
a challenge for nuclear theory.

Hardy and Towner have shown~\cite{HT05,TH08} that the calculated
corrections eliminate much of the considerable scatter present in the 
uncorrected $ft$ values, and the statistical consistency among
the corrected values is evidence that the corrections have been
reasonably computed. However, the importance of precisely testing 
the Standard Model stimulates us to undertake a reevaluation.
With this, we wish to start and stimulate further efforts 
to systematically improve ISB corrections, based on
an accurate understanding of ISB in nuclear forces~\cite{ISBref1,ISBref2}.

This paper is organized as follows. In Sect.~\ref{THproblem}, we show
that TH do not use the isospin operator to calculate $\delta_C$
(as mandated by the Standard Model). To examine potential consequences of
this, we review the TH treatment in Sect.~\ref{THdeltaC}.
A complete formalism is presented in Sect.~\ref{exact}, where we show
that there are no first order ISB corrections to the relevant nuclear
matrix elements, which is also true for the work of TH.
In Sect.~\ref{models}, we compare the TH treatment to
exact model results of increasing complexity, which can guide future
improvements. We conclude in Sect.~\ref{concl}.

\section{Towner and Hardy approach to ISB corrections}
\label{THproblem}

In nuclei, the matrix elements of weak vector interactions are not
modified by nuclear forces, except for corrections due to ISB effects.
Therefore, one has to evaluate the contributions from electromagnetic
and charge-dependent strong interactions to the Fermi matrix element
$M_F = \langle f | \tau_+ | i \rangle$ between the initial and final
states for superallowed $\beta$ decay, $| i \rangle$ and $| f \rangle$,
respectively. Here $\tau_+$ is the isospin raising operator.

Towner and Hardy~\cite{TH08} use a second quantization formulation
to write the Fermi matrix element as
\be
M_F = \sum_{\alpha, \beta} \langle f | a_{\alpha}^{\dag} a_{\beta} 
| i \rangle \langle \alpha | \tau_+ | \beta \rangle \,,
\label{MFq}
\ee
where $a_{\alpha}^{\dag}$ creates a neutron in state $\alpha$ and 
$a_{\beta}$ annihilates a proton in state $\beta$. Thus, the label 
$\alpha$ is used to denote neutron creation and annihilation operators,
while $\beta$ is used for those of the proton. This notation is 
different from the standard notation~\cite{DW},
in which $b_\alpha$ is used to
denote proton annihilation operators. 

The single-particle matrix element $\langle \alpha | \tau_+ | \beta 
\rangle$ is assumed to be given by the expression
\be
\langle \alpha | \tau_+ | \beta \rangle
= \delta_{\alpha, \beta} \int_0^{\infty}
R_{\alpha}^n(r) \, R_{\beta}^p(r) \, r^2 dr
\equiv \delta_{\alpha, \beta} \, r_{\alpha} \,,
\label{radi}
\ee
where $R_{\alpha}^n(r)$ and $R_{\beta}^p(r)$ are the neutron and proton
radial wave functions, respectively. The problem
is that the correct superallowed beta decay operator
in the Standard Model is the
plus component of the isospin operator. The operator in \eq{MFq} 
is not the isospin operator, because the states $| \alpha \rangle$ and
$| \beta \rangle$ are not the same. Instead, $\tau_+$ of 
\eq{MFq} is the plus component of the W-spin operator of 
MacDonald~\cite{Wspin}, which is reviewed in Ref.~\cite{DW}.
In addition, \eq{radi} assumes that the radial quantum numbers
of the states $\alpha$ and $\beta$ must be the same. This need not be so.
As a result, the Standard Model isospin commutation relations maintained
in the W-spin formalism are lost.

To obtain the commutation relations, we observe that Eqs.~(\ref{MFq}) 
and~(\ref{radi}) correspond to the second-quantized isospin operators
\bea
\tau_+ &=& \sum_{\alpha, \beta} \delta_{\alpha, \beta} \, r_\alpha \,
a^\dagger_\alpha a_\beta \,, \\
\tau_- &=& \tau_+^\dagger = \sum_{\alpha, \beta} \delta_{\alpha, \beta} \, 
r^*_{\alpha} \, a^\dagger_{\beta} a_{\alpha} \,,
\eea
so that
\be
[ \tau_+ , \tau_- ] = \sum_\alpha |r_\alpha|^2 \, 
a^\dagger_\alpha a_\alpha - \sum_\beta |r_\beta|^2 \, 
a^\dagger_\beta a_\beta \ne \tau_0 \,.
\ee
The Standard Model isospin commutation relations are violated if one 
uses the isospin operators of TH.

This formal problem motivates us to reevaluate the treatment of
ISB corrections, and to study whether there are potential corrections
to the extraction of $V_{ud}$. To this end, we review the details
of the TH procedure for $\delta_C$. Although Eqs.~(\ref{MFq}) and~(\ref{radi}) 
are not formally correct, they do account for the important correction:
the effects of the Coulomb interaction on the radial wave functions.
 
\section{TH treatment of $\delta_C$}
\label{THdeltaC}

Towner and Hardy~\cite{TH08} proceed by introducing into Eq.~(\ref{MFq})
a complete set of states for the $(A-1)$-particle system, $|\pi \rangle $,
which leads to
\be
M_F = \sum_{\alpha, \pi} \langle f | a_{\alpha}^{\dag} | \pi \rangle
\langle \pi | a_{\alpha} | i \rangle \, r_{\alpha}^{\pi} \,.
\label{MFpar}
\ee
The TH model thus allows for a dependence of the radial integrals
on the intermediate state $\pi$.

If isospin were an exact symmetry, the matrix elements of the creation
and annihilation operators would be related by hermiticity,
$\langle \pi | a_{\alpha} | i \rangle =
\langle f | a_{\alpha}^{\dag} | \pi \rangle^*$, and all radial integrals
would be unity. Hence the symmetry-limit matrix element in this model
is given by
\be
M_0 = \sum_{\alpha, \pi} | \langle f | a_{\alpha}^{\dag} | \pi \rangle 
|^2 \,.
\label{M0}
\ee
Towner and Hardy divide the contributions from ISB into two terms. First,
the hermiticity of the matrix elements of $a_{\alpha}$ and $a_{\alpha}^{\dag}$
will be broken, and second, the radial integrals will differ from unity.
Assuming both effects are small, TH calculate the resulting ISB
corrections as~\cite{TH08}
\be
\delta_C = \delta_{C1} + \delta_{C2} \,,
\label{dc1and2}
\ee
where in evaluating $\delta_{C1}$ all radial integrals are set to
unity but the matrix elements are not assumed to be related by 
hermiticity, and in evaluating $\delta_{C2}$ it is assumed that
$\langle \pi | a_{\alpha} | i \rangle =
\langle f | a_{\alpha}^{\dag} | \pi \rangle^*$ but $r_{\alpha}^{\pi}
\ne 1$. We will study whether this 
is a useful representation of $\delta_C$.
However, we emphasize that the separation into $\delta_{C1}$ and
$\delta_{C2}$ is a model-dependent concept, inspired by the shell 
model~\cite{TH02}.
For example, this division is clearly model dependent when $M_F$ 
is obtained from ab-initio calculations of the 
initial and final states, $| i \rangle$ and $| f \rangle$. In
addition, we demonstrate below that this is not possible rigorously
for schematic models.

\subsection{Radial overlap correction $\delta_{C2}$}

Towner and Hardy find that the radial correction,
$\delta_{C2}$, is the larger of their two model 
corrections~\cite{TH02,OB95,TH08}. The Fermi matrix element
relevant for $\delta_{C2}$ is given by
\bea
M_F & = & \sum_{\alpha, \pi} | \langle f | a_{\alpha}^{\dag} | 
\pi \rangle |^2 \, r_{\alpha}^{\pi} \,, \nonumber \\
& = & M_0 \left( 1 - \frac{1}{M_0}
\sum_{\alpha, \pi} 
|\langle f | a_{\alpha}^{\dag} | \pi \rangle |^2 \, \Omega_{\alpha}^{\pi} 
\right) \,,
\eea
where $\Omega_{\alpha}^{\pi} = (1 - r_{\alpha}^{\pi} )$ is a 
radial-mismatch factor. With the definition of the ISB 
correction factor in \eq{MF},
TH approximate $\delta_{C2}$ by
\be
\delta_{C2} \approx \frac{2}{M_0} \sum_{\alpha, \pi} 
|\langle f | a_{\alpha}^{\dag} | \pi \rangle |^2 \, \Omega_{\alpha}^{\pi} \,.
\label{dc2_1}
\ee
Consequently, large contributions
to $\delta_{C2}$ come with a large spectroscopic amplitude and a significant
radial mismatch.

In evaluating $\delta_{C2}$ of \eq{dc2_1}, TH use guidance from experiment.
Their results are based on shell model calculations of the spectroscopic
amplitudes, but limit the sums over orbitals $\alpha$ and intermediate 
states $\pi$ to those for which large spectroscopic factors have been 
observed in pick-up reactions. For $^{46}$V, TH~\cite{TH08} use 
this strategy to include two $sd$ core orbitals, $s_{1/2}$ and $d_{3/2}$,
in addition to the $f_{7/2}$ orbital of their earlier calculations.
Their new result for $\delta_{C2}$ is $0.58 \%$ (see Table~I in
Ref.~\cite{TH08}), which is almost a factor
two larger than the 2002 value~\cite{TH02}.

For the radial integrals,
TH use the strong constraint that the asymptotic forms of all radial 
functions must match experimentally measured neutron and proton separation
energies. In many cases, TH have to truncate the model space
to keep the calculations tractable. Their final values for $\delta_{C2}$
range between $0.17$ and $1.50 \%$, and increase with mass number (see
Table~II in Ref.~\cite{TH08}).

\subsection{Isospin-mixing correction $\delta_{C1}$}

The isospin-mixing correction $\delta_{C1}$ is obtained by setting all
radial integrals to unity, but including ISB corrections to the
matrix elements of the creation and annihilation operators,
$\langle f | a_{\alpha}^{\dag} | \pi \rangle^*$ and
$\langle \pi | a_{\alpha} | i \rangle$. These arise because the 
neutron-rich and proton-rich states are different. TH find that
calculations of $\delta_{C1}$ are very sensitive to the details 
of the shell-model computation, but try to reduce the model
dependence by using various experimental information~\cite{TH08}.

To obtain $\delta_{C1}$, TH use experimental single-particle energies
(on top of the core of the shell model calculation), which differ for
neutrons and protons. In addition, they include a two-body Coulomb 
interaction among the valence protons and increase all $T=1$
neutron-proton matrix elements (relative to the neutron-neutron
ones), so that the measured $b$ and $c$ coefficients of the
isobaric multiplet mass equation (IMME) are reproduced. Finally,
TH account for weak transitions that can occur to non-analog
$0^+$ states.  The adopted values for $\delta_{C1}$
range between $0.01$ and $0.35 \%$, and also increase with mass 
number (see Table~III in Ref.~\cite{TH08}).

\section{Exact formalism and theorems for ISB corrections}
\label{exact}

In this section, we present an exact formalism, independent of 
feasibility. We use this formalism to derive two theorems, which
show that there are no first-order ISB corrections to Fermi matrix
elements. This provides a perturbative expansion, which allows
for a simple estimate of $\delta_C$.

We use the correct isospin operator
\be
\tau_+ = \sum_\alpha a_\alpha^\dagger b_\alpha \,,
\label{true}
\ee
where $\alpha$ represents {\it any} single-particle basis, and 
$a^\dagger_\alpha$ creates neutrons and $b_\alpha$ annihilates protons
in state $\alpha$. The Fermi matrix element is then given by
\be
M_F = \langle f | \tau_+ | i \rangle \,, 
\label{truemf}
\ee
with $|i\rangle$ and $|f\rangle$ the exact initial and final 
eigenstates of the full Hamiltonian $H=H_0+V_C$, with energy
$E_i$ and $E_f$, respectively. Here $V_C$ denotes the sum of
{\it all} interactions that do not commute with the vector 
isospin operator ${\bf T} = \sum_{i=1}^A {\bm \tau}_i/2$,
\be
[H , {\bf T}] = [V_C , {\bf T}] \ne 0
\quad \text{and} \quad [H_0 , {\bf T}] = 0 \,.
\ee

We will use round bra and ket states to denote the eigenstates
of the isospin-symmetric part of the Hamiltonian, so $H_0 |n) = 
E_n^{(0)} |n)$. Obtaining the states $|n)$ requires a solution 
of the $A$-body problem.

The full initial eigenstate $|i\rangle$ can then be written as
\be
| i \rangle = \sqrt{Z_i} \, \biggl[ |i) + \frac{1}{E_i-\Lambda_i H
\Lambda_i}\, \Lambda_i V_c \, |i) \biggr] \,,
\label{initial}
\ee
with projector $\Lambda_i \equiv 1-|i)(i|$, or equivalently,
\be
|i\rangle = \sqrt{Z_i} \, |i) + \frac{1}{E_i-\Lambda_i H_0 \Lambda_i}
\, \Lambda_i V_c \, |i\rangle \,.
\ee
Similarly, the full final eigenstate $|f\rangle$ is given by
\be
| f \rangle = \sqrt{Z_f} \, \biggl[ |f) + \frac{1}{E_f-\Lambda_f H
\Lambda_f}\, \Lambda_f V_c \, |f) \biggr] \,,
\label{final}
\ee
with $\Lambda_f = 1-|f)(f|$. The factors $Z_i$ and $Z_f$ are taken 
to be real and ensure that the full eigenstates $|i\rangle$ and 
$|f\rangle$ are normalized. As a result, it follows (due to the
projection operators) that the deviations of $Z_i$ and $Z_f$ from
unity start at second order in $V_c$.

We now evaluate \eq{truemf} between the exact eigenstates given 
by Eqs.~(\ref{initial}) and~(\ref{final}). With $(f| \, \tau_+ 
\Lambda_i = 0$ and $\Lambda_f \tau_+ \, |i)=0$, we obtain
\begin{widetext}
\be
M_F = \sqrt{Z_i Z_f} \, \biggl[ M_0 + (f| \, V_c \Lambda_f \,
\frac{1}{E_f-\Lambda_f H \Lambda_f} \, \tau_+ \,
\frac{1}{E_i-\Lambda_i H \Lambda_i} \, \Lambda_i V_c \, |i) \,
\biggr] \,,
\ee
\end{widetext}
where $M_0 = ( f | \tau_+ | i )$. Since $Z_{i,f} = 1 + 
{\mathcal O}(V_C^2)$, it follows that ISB contributions start
at second order. This is our first theorem and demonstrates that 
there are no first-order ISB corrections to $M_F$.

We obtain a simpler form by expanding in the {\it difference}
of the charge-dependent interactions $\Delta V_C$
between the initial proton-rich and final neutron-rich states.
Hence, $\Delta V_C$ includes all charge-dependent interactions
of the extra proton with the other nucleons in the initial state. 
In this case, we have
\be
|f\rangle = |f) \quad \text{and} \quad
|i\rangle =
\sqrt{Z} \, |i) + \frac{1}{E_i-\Lambda_i \widetilde{H}_0 \Lambda_i} \,
\Lambda_i \Delta V_C \, |i\rangle \,,
\ee
where the first expression defines $\Delta V_C$ and $\widetilde{H}_0$
includes the effects of $V_C$ common to the 
initial and final states, for example the Coulomb interactions in
the core.

In this case, the final state is an eigenstate of 
$\widetilde{H}_0$ and obeys $\langle f| \, \tau_+ \Lambda_i = 0$.
As a result, it follows that
\be
M_F = \sqrt{Z} \, M_0 \,.
\label{th2}
\ee
This is our second theorem. As already shown, there are no first-order
ISB corrections to Fermi matrix elements, and in this case $\delta_C 
= 1 - Z$ has a straightforward perturbative expansion in $\Delta V_C$,
starting at second order:
\begin{widetext}
\be
\delta_C = ( i| \, \Delta V_C \Lambda_i \, \biggl( \frac{1}{E_i - \Lambda_i 
\widetilde{H}_0 \Lambda_i} \biggr)^2 \Lambda_i \Delta V_C \, |i)
+ 2 \: {\rm Re} \biggl[ (i| \, \Delta V_C  \Lambda_i \, \biggl( \frac{1}{E_i-
\Lambda_i\widetilde{H}_0\Lambda_i} \biggr)^2 \Lambda_i \Delta V_C \, 
\frac{1}{E_i-\Lambda_i\widetilde{H}_0\Lambda_i} \, \Lambda_i \Delta V_C \,
|i) \biggr] + \ldots \,,
\label{znorm}
\ee
\end{widetext}
which follows from the normalization condition $\langle i | i \rangle = 1$.
To second order in $\Delta V_C$, the full energy $E_i$ can be taken as
the energy $\widetilde{E}_i^{(0)}$ of $\widetilde{H}_0$.
Examining the third-order term of \eq{znorm}~\footnote{The full
evaluation of third-order corrections requires a calculation of 
$E_i$ to first order in $\Delta V_C$.}
also shows that it is impossible to separate $\delta_C$ into 
two distinct terms. This is because it is not possible to 
distinguish whether the  middle $\Delta V_C$ is part of a 
correction to an intermediate state $|\pi \rangle$ or to the
initial state $|i \rangle$. Finally, we note that the two
theorems are more general versions of the theorem of 
Behrends-Sirlin~\cite{BS}
for CVC in nucleons and of Ademollo-Gatto~\cite{AG} for weak decays of kaons.

\subsection{Simple estimate of  $\delta_C$}

As an illustration of the above formalism, we calculate $\delta_C$ for
the case of a single particle outside an inert core of charge $Ze$, assuming 
harmonic-oscillator single-particle wave functions with
oscillator frequency $\omega \approx 39 \mev \, A^{-1/3}$.
The nuclear Coulomb potential arises from the convolution of 
$Z e^2/(4 \pi |\bfr-\bfr'|)$ with the charge density $\rho_C(r')$. If we 
take the latter to be a constant within $r \leqslant R$,
the one-body Coulomb potential
takes the simple form:
\bea
\Delta V_C(r) = \frac{Z e^2}{4 \pi R} \biggl[ \Theta(R-r) \biggl( \frac{3}{2}
- \frac{r^2}{2 R^2} \biggr) +\Theta(r-R) \, \frac{R}{r} \biggr] \,.
\label{VCsimp}\eea
With $R = 1.1 \fm \, A^{1/3}$, we have $R \omega = 0.22$, independent
of $A$, and therefore the correction scales 
as $\delta_C \sim Z^2$. To make an estimate,
we take the state $|i)$ to be in the single-particle orbit with
radial quantum number $n=0$ and angular momentum $l$. Using
\eq{znorm}, we find
\begin{widetext}
\be
\delta_C(l) = \frac{Z^2 e^4}{4 (4 \pi)^2 R^2 \omega^2} \sum\limits_{n > 0}
\, \frac{1}{n^2} \: \biggl\{ \int r^2 dr \, R_{0l}(r)
\biggl[ \Theta(R-r) \biggl( \frac{3}{2}
- \frac{r^2}{2 R^2} \biggr) +\Theta(r-R) \, \frac{R}{r} \biggr]
R_{nl}(r) \biggr\}^2 \,.
\ee
\end{widetext}
We calculate the summation numerically taking $R$ equal to the oscillator
length. This leads to
\bea
&\delta_C(0) = 0.0020 \% \, Z^2 \:\: &
\delta_C(1) = 0.0013 \% \, Z^2 \nonumber\\[2mm]
&\delta_C(2) = 0.00071 \% \, Z^2 \:\: &
\delta_C(3) = 0.00043 \% \, Z^2 \,.
\eea
For a $Z=20$ core and $l=3$, we find $\delta_C(3) = 0.17 \%$ and thus $\delta_C 
= 3.33 \times 
\delta_C(3) = 0.57 \%$, where the factor of $3.33$ arises from the 
Macfarlane-French sum rule~\cite{MF} for the three protons in the $f_{7/2}$
orbit. This result is in qualitative agreement with the TH contribution of 
$0.45 \%$ (see Table~I for $^{46}V$ in Ref.~\cite{TH08}).
This indicates that \eq{th2} could be a useful starting point for 
realistic calculations.

\section{Schematic models}
\label{models}

Next we present exact evaluations of the Fermi matrix element for
schematic models of increasing complexity, and compare our results
to the treatment of TH.

\subsection {One-body problem}

Consider starting from the exact formalism. We can derive an effective
single-particle potential $U + U_C$, where $U_C$ accounts for
charge-dependent effects and acts only on protons. The
single-particle potential is introduced to minimize the effects
of residual interactions
\be
\Delta V = V + V_C - (U + U_C) \,.
\ee
Then the Hamiltonian is given by
\be
H = T + U + U_C + \Delta V = H_{\rm sp} + \Delta V \,.
\ee
In the simplest case, we assume that the one-body Hamiltonian 
$H_{\rm sp}$ is dominant. 
 Thus we take the initial and final states 
to consist of a single nucleon outside an inert core $|0\rangle$.
The core and nucleon have quantum numbers so that the coupled
state is $0^+$ with $T=1$.

The one-body basis states can be taken as eigenstates of the
full single-particle Hamiltonian $T + U + U_C$, which we denote
by $|\alpha\rangle$, or by the eigenstates $|\widetilde{\alpha}
\rangle$ of the isospin-symmetric part $T + U$.
Here $U_C$ is the difference between the proton and neutron potentials,
which corresponds to $U_C \equiv U_p-U_n+V_C$ in the TH notation.
The parent and
daughter states are then given by
\be
|i\rangle = b^\dagger_\alpha |0\rangle \quad \text{and}
\quad |f\rangle = a^\dagger_{\widetilde{\alpha}} |0\rangle \,.
\ee

It is convenient to express the isospin raising operator 
$\tau_+$ in a mixed representation. The creation operators of 
the two bases are related by
\be
a^\dagger_{\alpha} = \sum_{\widetilde{\alpha}} 
a^\dagger_{\widetilde{\alpha}} \, 
\langle\widetilde{\alpha}|\alpha\rangle \,,
\ee
and therefore the isospin operator of \eq{true} reads
\be
\tau_+ = \sum_{\alpha,\widetilde{\alpha}}
a^\dagger_{\widetilde{\alpha}} \,
\langle\widetilde{\alpha}|\alpha\rangle \, b_\alpha \,.
\ee
This equation leads to an expression for $M_F$ that is very similar
to \eq{radi} of TH with the important difference that the states 
$\alpha$ and $\widetilde{\alpha}$ need not have the same radial quantum
numbers:
\be
M_F/M_0 = \int r^2 dr \, R^*_{\widetilde{\alpha}}(r) \, R_\alpha(r) \,.
\label{exact1}
\ee

\subsection{One-body problem with a single core excitation}

The simplest generalization of the previous problem is to allow 
the core to have two states, a ground state and excited state. 
Then the exact eigenstate is a two component wave function, where
the upper component represents the single particle plus unexcited
core, and the lower component has the core in the excited state.
The core excitation need not have angular momentum $J=0$, but the
coupled state is $0^+$ with $T=1$. In this case, the Hamiltonian
is given as a two-by-two matrix:
\be
H_0 = \left( \begin{array}{cc} H_{\rm sp} & \Delta V \\
\Delta V  &  H_{\rm sp} \end{array} \right) \,,
\ee
where the second, lower component has a higher single-particle 
energy than the upper component. The eigenfunctions are
given by two-component ``spinor'' wave functions, for example for
the initial state:
\be
\langle \bfr | i \rangle = \left( \begin {array}{c}
\alpha_i \, U_i(\bfr) \\
\beta_i \, L_i(\bfr) \end{array} \right) \,,
\ee
with normalizations given by
\be
\int d\bfr \, |U_i(\bfr)|^2 = \int d\bfr \, |L_i(\bfr)|^2 = 1
\quad \text{and} \quad \alpha_i^2 + \beta_i^2 = 1 \,,
\label{norm}
\ee
and where we have taken $\alpha, \beta$ to be real for simplicity.

The presence of the charge-dependent interaction $U_C$ (in $H_{\rm sp}$)
and of $V_C$ (in $\Delta V$) causes the initial and final state values
of $\alpha, \beta$ and their radial wave functions to differ. In this
model, the single-particle wave functions for $i$ and $f$ represent
directly the single proton and neutron. The exact value of $M_F$ is
thus given by:
\begin{multline}
M_F/M_0 = \alpha_f \, \alpha_i \int d\bfr \, U^*_f(\bfr) \, U_i(\bfr) \\
+ \beta_f \, \beta_i \int d\bfr \, L^*_f(\bfr) \, L_i(\bfr) \,,
\label{exact2}
\end{multline}
since the core and its excitation are orthogonal. This may be rewritten
as:
\begin{multline}
M_F/M_0 - 1 = - \alpha_i^2 \, \Omega^{(1)} - \beta_i^2 \, \Omega^{(2)} \\[2mm]
+ (\alpha_f-\alpha_i) \, \alpha_i \, (1-\Omega^{(1)})
+ (\beta_f-\beta_i) \, \beta_i \, (1-\Omega^{(2)}) \,,
\label{exact2b}
\end{multline}
where in the TH notation
\bea
\Omega^{(1)} &=& 1 - \int d\bfr \, U^*_f(\bfr) \, U_i(\bfr) \,, \\
\Omega^{(2)} &=& 1 - \int d\bfr \, L^*_f(\bfr) \, L_i(\bfr) \,.
\eea
We next use the strategy of TH to evaluate this two-state core model. 
The states $|\pi\rangle$ consist of the ground state core and its 
excitation labeled by $1, 2$. For each of these, there is only 
one value for the single-particle index $\alpha$. 
Therefore, the TH result for this model reads
\bea
M_F^{\text{TH}}/M_0 - 1 &\approx& - \alpha_i^2 \, \Omega^{(1)} - \beta_i^2 \,
\Omega^{(2)} \nonumber \\[2mm]
&+& (\alpha_f-\alpha_i) \, \alpha_i + (\beta_f-\beta_i) \, \beta_i \,.
\label{mfth}
\eea
The contributions on the first line of Eq.~(\ref{mfth}) correspond to
$\delta_{C2}$ and the terms on the second line to $\delta_{C1}$.

In comparison with the exact result, we observe that 
TH neglect terms of order $(\alpha_f-\alpha_i) \Omega^{(1)}$.
The relevant radial integrals are of infinite order
in $U_C$, so that setting them to unity in evaluating the
second line of \eq{mfth} may be significant relative to the
required accuracy, in particular if the neutron and
proton separation energies are very different. In addition, 
this schematic model indicates that the normalization conventions
of \eq{norm} are just a choice, so that the separation into the two 
terms $\delta_{C1}$, $\delta_{C2}$ seems rather arbitrary. It 
is just as reasonable to use the product $\alpha_i U_i(\bfr)$
as the upper component $\widetilde{U}_i(\bfr)$ and $\beta_i L_i(\bfr)$
as the lower component $\widetilde{L}_i(\bfr)$ with the normalization
$\int d\bfr \, \big(|\widetilde{U}_i(\bfr)|^2 + |\widetilde{L}_i(\bfr)|^2
\bigr)$. For that convention, the factors $\alpha_{i,f}, \beta_{i,f}$ would
disappear from the formalism, and the separation of $\delta_C$ into
$\delta_{C1}$, $\delta_{C2}$ would neither be necessary nor possible.

\subsubsection{Evaluation using simple interactions}

Let the state $|f\rangle$ be governed by a harmonic oscillator Hamiltonian 
$H_0$,
\be
H_0 = \frac{p^2}{2 m} + \frac{1}{2} m \omega^2 r^2 \,,
\ee
and the state $|i\rangle$ by the Hamiltonian $H$, with
\be
H=H_0 + V_C \,.
\ee
We use the inner form of the Coulomb potential, \eq{VCsimp},
to obtain simple expressions that show the order of various terms,
\be
V_C(r) = - \frac{Z e^2}{4 \pi R} \, \frac{r^2}{2 R^2} \,.
\ee
This is a qualitative approximation that has been traditionally used
to assess the size of various effects~\cite{DW}.

In this case, the Coulomb interaction shifts the
square of the oscillator frequency from $\omega^2$ to $\omega^2 
(1-\delta)$, where the shift $\delta$ is given by
\be
\delta = \frac{Z e^2}{4 \pi m \omega^2 R^3} \,.
\ee
We will consider the two lowest states with $n=0, 1$ and angular
momentum $l=0$ to study the effects of configuration mixing.

Then for the final state single-particle basis, we have
\bea
(\bfr|0) &=& \frac{1}{(\pi b^2)^{3/4}} \, e^{-r^2/2 b^2} \,, \\[1mm]
(\bfr|1) &=& \sqrt{\frac{3}{2}} \biggl( 1 - \frac{2 r^2}{3 b^2} \biggr)
\frac{1}{(\pi b^2)^{3/4}} \, e^{-r^2/2 b^2} \,,
\eea
with oscillator length $b = (m \omega)^{-1/2}$. The initial state 
wave functions $(\bfr|\widetilde{0})$ and $(\bfr|\widetilde{1})$
have the same form, but with oscillator length
\be
\widetilde{b^2} = \frac{b^2}{1 - \delta/2} \,.
\ee
With this, we find the radial overlaps:
\bea
(0|\widetilde{0}) &=& \biggl[ \frac{1-\delta/2}{(1-\delta/4)^2} \biggr]^{3/4}
= 1 - \frac{3 \, \delta^2}{64} + {\mathcal O}(\delta^3) \,, \\[2mm]
(1|\widetilde{1}) &=& \frac{(1-\delta/2)^{3/4} \, (1-\delta/2-3\delta^2/32)
}{(1-\delta/4)^{7/2}} \,, \\[2mm]
(0|\widetilde{1}) &=& \frac{\sqrt{3} \, \delta \, (1-\delta/2)^{3/4}}{
(2-\delta/2)^{5/2}} \,, \label{01} \\[2mm]
(1|\widetilde{0}) &=& - (0|\widetilde{1}) \,.
\eea

Now consider configuration mixing between the states 0 and 1 by strong
interactions. As as a result, we have
\bea
|f\rangle &=& \alpha \, |0) + \beta \, |1) \,,\\[2mm]
|i\rangle &=& \widetilde{\alpha} \, |\widetilde{0})
+ \widetilde{\beta} \, |\widetilde{1}) \,,
\eea
where we take $\alpha,\beta, \widetilde{\alpha}$ and $\widetilde{\beta}$
to be real for simplicity, 
and the quantities $\widetilde{\alpha}-\alpha$ and 
$\widetilde{\beta}-\beta$ are of order $\delta^2$.

The Fermi matrix element is then given by
\be
M_F/M_0 = \alpha \widetilde{\alpha} \, (0|\widetilde{0})
+ \beta \widetilde{\beta} \, (1|\widetilde{1})
+ [\alpha \widetilde{\beta} - \beta \widetilde{\alpha}] 
\, (0|\widetilde{1}) \,.
\label{xmple}
\ee
The first two terms are equal to $1 + {\mathcal O}(\delta^2)$. Since
$\widetilde{\alpha} = \alpha + {\mathcal O}(\delta^2)$ and similarly
for $\widetilde{\beta}$, the leading first-order part of 
$(0|\widetilde{1})$ (see Eq.~(\ref{01})) 
thus cancels exactly, and the last two terms of
\eq{xmple} start at order $\delta^3$. This validates that 
there are no first-order ISB corrections to $M_F$,
and shows that certain approximations can violate our theorems.

\subsection{Two interacting nucleons outside an inert core}

Next we assume an inert core and two interacting nucleons with total angular
momentum $J = 0$. The initial state wave function can have components 
spread over
different single-particle configurations with radial, orbital and total
quantum numbers $n, l, j$:
\begin{multline}
|i\rangle =\sum\limits_{n_1,n_2,l,j,m} \langle j,m,j,-m | 0,0 \rangle \\
\times A_i^{n_1,n_2,l,j} \, b^\dagger_{n_1 l j,m}
\, b^\dagger_{n_2 l j,-m} \, |0\rangle \,.
\end{multline}
For clarity, we have taken two protons $(b^\dagger)^2$ on top of a $0^+$
core, coupled to $J=0$, $M_J=0$. A similar expression for the final state
involves the amplitudes $A_f^{n_1,n_2,l,j}$, which differ due to the
effects of the charge-dependent interactions. In this way,
two-nucleon correlations are incorporated in a limiting case of the 
formalism of Sect.~\ref{exact}.

The exact expression for $M_F$ is then given by
\begin{multline}
M_F/M_0 =\sum\limits_{n_1,n_1',n_2,l,j} A_f^{n_1,n_2,l,j\,*} \, A_i^{n_1',n_2,l,j} \\
\times \int r^2 dr \, R_f^{n_1' l\,*}(r) R_i^{n_1 l}(r) \,,
\label{exact3}
\end{multline}
with radial wave functions $R_{i,f}$. For example, $n_1$ could correspond to
states in the shell-model valence space and $n_1'$ to a high-lying
shell due to strong interactions.

The TH approximation for $M_F$ would be
\begin{multline}
M_F^{\rm TH}/M_0 \approx \sum\limits_{n_1,n_2,l,j}^{\text{model space}} \biggl[ \,
| A_i^{n_1,n_2,l,j} |^2 \, \bigl( 1 - \Omega^{(n_1 l j)} \bigr) \\
+\bigl( A_f^{n_1,n_2,l,j\,*} - A_i^{n_1,n_2,l,j\,*} \bigr) \,
A_i^{n_1,n_2,l,j} \, \biggr] \,.
\label{TH3}
\end{multline}
We compare the exact result of \eq{exact3} with the TH approximation \eq{TH3}:
\begin{enumerate}
\item We find corrections to the radial overlaps, because the
quantum numbers $n_1$ and $n_1'$ need not be equal.
\item The exact result mixes in higher-lying configurations that are not
within the TH model space. To incorporate ISB effects due to higher-lying
states, one needs to evaluate their contributions to charge-dependent 
effective interactions. In particular, an interesting topic for future 
study is the renormalization from long-range Coulomb effects.
\item As in the previous models, the radial integrals of \eq{exact3} are 
of infinite order in $V_C$, so that setting them to unity in evaluating
$\delta_{C2}$ might not be very accurate.
\end{enumerate}

\subsection{Two nucleons with a single core excitation}

This model combines those of the two previous subsections. The core 
can be excited so that the two interacting nucleons are outside a 
core in its ground or excited state. Using the previously adopted
notation, the exact value of $M_F$ reads
\begin{align}
& M_F/M_0 = \sum\limits_{n_1,n_1',n_2,l,j} \nonumber \\
&\times \biggl[ \, A_f^{n_1,n_2,l,j\,*} \, A_i^{n_1',n_2,l,j} 
\int r^2 dr \, U_f^{n_1' l\,*}(r) U_i^{n_1 l}(r) \nonumber \\[2mm]
&+ B_f^{n_1,n_2,l,j\,*} \, B_i^{n_1',n_2,l,j}
\int r^2 dr \, L_f^{n_1' l\,*}(r) L_i^{n_1 l}(r) \, \biggr] \,,
\label{exact4}
\end{align}
and the TH approximation would be
\begin{align}
&M_F^{\rm TH}/M_0 \approx \sum\limits_{n_1,n_2,l,j}^{\text{model space}} \biggl[ \,
| A_i^{n_1,n_2,l,j} |^2 \, \bigl( 1 - \Omega_1^{(n_1 l j)} \bigr) \nonumber \\[1mm]
&+\bigl( A_f^{n_1,n_2,l,j\,*} - A_i^{n_1,n_2,l,j\,*} \bigr) \,
A_i^{n_1,n_2,l,j} \nonumber \\[2mm]
&+ | B_i^{n_1,n_2,l,j} |^2 \, \bigl( 1 - \Omega_2^{(n_1 l j)} \bigr) 
\nonumber \\[2mm]
&+\bigl( B_f^{n_1,n_2,l,j\,*} - B_i^{n_1,n_2,l,j\,*} \bigr) \,
B_i^{n_1,n_2,l,j} \, \biggr] \,.
\label{TH4}
\end{align}

\section{Conclusions}
\label{concl}

We have studied the formalism to include ISB corrections to Fermi 
matrix elements, motivated by the recent experimental achievements 
on $0^{+} \rightarrow 0^{+}$ nuclear $\beta$ decay and by the work
of Towner and Hardy~\cite{TH08}. This is a key challenge for nuclear
theory and pivotal for extracting the up-down CKM matrix element,
which provides precision tests of the Standard Model.

We have shown that TH do not use the isospin operator of the Standard
Model to calculate ISB corrections. It is also true that their separation
$\delta_{C} = \delta_{C1}+\delta_{C2}$ is model dependent~\cite{TH02}. 
Using a complete formalism, we derived two theorems, which demonstrate 
there are no first-order ISB corrections to Fermi matrix elements. 

Towner and Hardy correctly
include the leading part of the Coulomb effects on the radial
wave functions.
We have found corrections to the TH treatment, and contrasted these
to exact results in schematic models of increasing complexity.
One of the differences is that the radial overlaps need not have
the same radial quantum numbers (as assumed in TH). 
This mixing has also been pointed out in density-functional 
based calculations of ISB corrections~\cite{DH}. In addition, 
significant to the required accuracy, there may be contributions
from higher-lying configurations that are outside the model 
space. This requires a careful inclusion into 
charge-dependent effective interactions, where our accurate
understanding of isospin-symmetry breaking in nuclear forces
can be very helpful, and a careful study of truncation effects,
where modern methods can lead to improvements. 

Numerical evaluations using the formalism presented
here are needed as a next step. We hope that our work
stimulates further efforts to make systematic improvements
to the important problem of ISB corrections to superallowed
transitions.

\acknowledgments

We thank Dick Furnstahl, Paul Garrett, Geoff Grinyer,
John Hardy and Ian Towner for useful discussions.
This work was supported in part by the US Department of Energy 
under Grant No.~DE--FG02--97ER41014 and the Natural Sciences and 
Engineering Research Council of Canada (NSERC). TRIUMF receives federal
funding via a contribution agreement through the National Research 
Council of Canada.


\begin{thebibliography}{99}
\bibitem{HT05} J.C.\ Hardy and I.S.\ Towner, \prc {\bf 71}, 055501 (2005).
\bibitem{HT05a} J.C.\ Hardy and I.S.\ Towner, \prl {\bf 94}, 092502 (2005).
\bibitem{Ha07} J.C.\ Hardy, hep-ph/0703165.
\bibitem{TH02} I.S.\ Towner and J.C.\ Hardy, \prc {\bf 66}, 035501 (2002).
\bibitem{OB95} W.E.\ Ormand and B.A.\ Brown, \prc {\bf 52}, 2455 (1995).
\bibitem{TH08} I.S.\ Towner and J.C.\ Hardy, \prc {\bf 77}, 025501 (2008).
\bibitem{Sa05} G.\ Savard {\it et al.}, \prl {\bf 95}, 102501 (2005).
\bibitem{Er06b} T.\ Eronen {\it et al.}, \prl {\bf 97}, 232501 (2006).
\bibitem{ISBref1} G.A.\ Miller, B.M.K.\ Nefkens and I.\ Slaus, Phys.\ Rept.\ 
{\bf 194}, 1 (1990).
\bibitem{ISBref2} G.A.\ Miller, A.K.\ Opper and E.J.\ Stephenson, Ann.\ 
Rev.\ Nucl.\ Part.\ Sci.\ {\bf 56}, 253 (2006).
\bibitem{DW} D.H.\ Wilkinson, Ed., ``Isospin in Nuclear Physics'', North 
Holland, 1969.
\bibitem{Wspin} W.M.\ MacDonald, in ``Isobaric Spin in Nuclear Physics'', 
Ed.~by J.D.\ Fox and D.\ Robson, Academic Press, 1966.
\bibitem{BS} R.E.\ Behrends and A.\ Sirlin, \prl {\bf 4}, 186 (1960).
\bibitem{AG} M.\ Ademollo and R.\ Gatto, \prl {\bf 13}, 264 (1964).
\bibitem{MF} J.B.\ French and M.H.\ Macfarlane, Nucl. Phys. {\bf 26}, 168 
(1961). 
\bibitem{DH} J.\ Dobaczewski and I.\ Hamamoto, Phys.\ Lett.\ B {\bf 345},
181 (1995).
\end{thebibliography}
\end{document}